\documentclass[axodraw,12pt]{article}
\usepackage{axodraw}
\usepackage{paper2e}
\usepackage{mydefs2e}
\usepackage{graphicx}
\usepackage{hyperref}

\renewcommand{\Box}{\,\raisebox{-.45pt}{\drawsquare{7}{0.6}}\,}

\renewcommand{\d}{\partial}
\newcommand{\dd}{\raisebox{1.2pt}{$\stackrel{\raisebox{-1pt}%
{$\scriptscriptstyle\leftrightarrow$}}{\d}$}}

\def\bea{\begin{eqnarray}}
\def\eea{\end{eqnarray}}

\begin{document}

\begin{titlepage}
\renewcommand{\thefootnote}{\arabic{footnote}}

\preprint{UMD-PP-03-008}

\title{Supergravity Loop Contributions\\\medskip
to Brane World Supersymmetry Breaking}

\author{I.L. Buchbinder$^{\,\rm a}$\footnote{{\tt joseph@tspu.edu.ru}},
$~~$S. James Gates, Jr.$^{\,\rm b, c}$\footnote{{\tt gatess@wam.umd.edu}},
$~~$Hock-Seng Goh$^{\,\rm c}$\footnote{{\tt hsgoh@physics.umd.edu}},\\
W. D. Linch III$^{\,\rm b, c}$\footnote{{\tt
ldw@physics.umd.edu}}, $~~$Markus A. Luty$^{\,\rm
c}$\footnote{{\tt luty@umd.edu}},
 $~~$Siew-Phang Ng$^{\,\rm
c}$\footnote{{\tt spng@physics.umd.edu}},
$~~$J. Phillips$^{\,\rm
b, c}$\footnote{{\tt ferrigno@physics.umd.edu}}}

\address{$^{\rm a}$Department of Theoretical Physics, Tomsk State Pedagogical University\\
634041 Tomsk, Russia}

\address{$^{\rm b}$Center for String and Particle Theory,
University of Maryland\\
College Park, Maryland 20742, USA}

\address{$^{\rm c}$Department of Physics, University of Maryland\\
College Park, Maryland 20742, USA}

\begin{abstract}
We compute the supergravity loop contributions to the visible
sector scalar masses in the simplest 5D `brane-world' model.
Supersymmetry is assumed to be broken away from the visible brane
and the contributions are UV finite due to 5D locality.
We perform the calculation with $\scr{N} = 1$ supergraphs,
using a formulation of 5D supergravity in terms of $\scr{N} = 1$
superfields.
We compute contributions to the 4D effective action that
determine the visible scalar masses, and we find that the
mass-squared terms are negative.
\end{abstract}

\end{titlepage}
\section{Introduction}

\noindent
In this paper, we study supersymmetry (SUSY) breaking in the simplest
5D `brane world' scenario.
In brane world scenarios, some or all of the visible sector fields are assumed
to be localized on a brane, and SUSY is broken away from the
visible brane.  In this case, bulk fields transmit the message of SUSY breaking to the
visible sector.  We consider the minimal case where the bulk fields are the 5D supergravity (SUGRA) multiplet.  Thus, supergravity plays the role of the messenger for
SUSY breaking.  Previously, \Ref{RS} showed that the leading contribution to
visible sector SUSY breaking for large radius, comes from
anomaly-mediated SUSY breaking (see also \Ref{GLMR}).
If the visible sector consists only of the minimal supersymmetric
standard model, the slepton mass-squared terms are negative.
Thus, for these brane-world models to be realistic we require other
contributions to SUSY breaking in the visible sector.
With the hope of getting positive mass-squared terms, we will calculate the
leading contributions to SUSY breaking by SUGRA loops.

The simplest 5D brane-world scenario can be described as follows.
The 5D space-time is flat and compactified on an $S^1 / Z_2$ orbifold.
There is one 3-brane at each of the $Z_2$ fixed points.
These 3-branes can be regarded as the boundaries of
the extra dimension of length $\ell = \pi r$,
where $r$ is the radius of the $S^1$.
We assume that SUSY is broken by the vacuum expectation value of
a chiral superfield $X$ localized on the hidden brane.
The visible chiral superfields $Q$ are assumed to be
localized on the other brane.
In this 5D effective theory, contact terms between $Q$ and $X$ are forbidden by 5D locality.%
\footnote{In a more fundamental theory with additional states
with masses $M \gg 1/r$, contact terms between $Q$ and $X$ will be suppressed by
$e^{-M r}$.}
The effects of SUGRA mediated SUSY breaking can be analyzed
systematically using the 4D effective Lagrangian that describes
the physics below the compactification scale $1/r$. The effective
theory contains the chiral superfields $Q$ and $X$, the 4D SUGRA
multiplet, and the chiral radion multiplet
\beq
T = \pi r + \cdots + \theta^2 F_T.
\eeq
Expanding the 4D effective action in $Q$ and $X$,
the leading terms involving $Q$ that cannot be
forbidden by symmetries are
\beq[c12]
\De \scr{L}_{\rm 4,eff}
= \myint d^4\theta\left[ c_1(T) Q^\dagger Q + c_2(T) X^\dagger X
Q^\dagger Q + \cdots \right].
\eeq
At tree level, $c_1$ is independent of $T$, and
\Ref{LS1} showed that
$c_2$ vanishes.  Therefore, we must
consider loop effects.

At 1-loop level, there are contributions to $c_1$ from the diagrams in Fig.~1.
These contributions are of order
\beq
c_1 \sim \frac{1}{M_5^3 (T + T^\dagger)^3}.
\eeq
The dependence on $T$ is fixed by dimensional
analysis and the observation that $c_1$ cannot depend on the fifth
component of the graviphoton of 5D SUGRA, which is contained in
$T-T^\dag$ \cite{LS1}.
Loop corrections to $c_1$ are finite
because they are sensitive to the size of the extra dimension,
while all divergent effects are local.
These corrections can give important contributions to the
scalar masses of $Q$ if $\avg{F_T} \ne 0$:
\beq[sweet]
\De m^2_Q = -3 \avg{c_1} \left| \left\langle
\frac{F_T}{T} \right\rangle \right|^2.
\eeq
Here we have
neglected 1-loop operators of the form
\beq
\De\scr{L}_{\rm 4,eff}
\sim \myint d^4\theta\, \frac{| D^2 T |^2} {M_5^3 (T + T^\dagger)^2}
Q^\dagger Q,
\eeq
which give contributions to the scalar masses
proportional to $\avg{F_T}^4$.
Thus, the contribution from $c_1$ in \Eq{sweet} dominates only if $\avg{F_T} \ll 1$.
A nonzero value for $\avg{F_T}$ is equivalent to the Scherk--Schwarz
\cite{SS} mechanism for SUSY breaking, as discussed in
\Ref{SSequiv}.
The SUGRA loop effect proportional to $c_1$
was computed in \Ref{GR} using the
off-shell formulation of supergravity due to Zucker \cite{Zucker}.
It was found that the resulting scalar mass-squared terms are
negative.

There are 1-loop
contributions to $c_2$ from the diagrams in Fig 2.
These diagrams are UV finite because the loop cannot shrink to zero size.
By dimensional analysis, these give
\beq
c_2 \sim \frac{1}{M_5^6 (T + T^\dagger)^4}.
\eeq
This is suppressed by extra powers of $M_5$ compared to $c_1$.
This contribution may be important if $\avg{F_T}$ is sufficiently small.
In this case, it gives a contribution to the $Q$ scalar mass
\beq
\De m^2_{Q} = - \avg{c_2} |\avg{F_X}|^2.
\eeq

Although $c_1$ is known in the literature, $c_2$ has never been calculated.
In this paper, we will present explicit calculations of both $c_1$ and $c_2$.
We perform quantum computations using supergraphs (see e.g. \cite{GGRS}, \cite{BK}) applied to the formulation of 5D SUGRA in $\scr{N} = 1$ superspace developed in \Ref{LLP}.
This formalism has several advantages over component calculations.
First, higher powers of Dirac delta functions from brane-bulk interactions
do not arise in this formulation.
Higher powers of Dirac delta functions occur only after integrating
out auxiliary fields \cite{MP}, and therefore are absent in
supergraph calculations.
Furthermore, the gauge can be fixed so that the superspace supergravity
propagator has the following trivial tensor structure:
\beq[boney]
\avg{V_m V_n} \sim
\frac{\eta_{mn}}{\Box_5},
\eeq
where $m, n = 0, \ldots 3$ are 4D Lorentz indices and $V_m$
is the SUGRA superfield prepotential.
The simple form of this propagator makes quantum calculations straightforward.
Another advantage of this approach is that we only need to calculate
five super Feynman graphs.
In a direct component formulation, this number would grow
by an order of magnitude.

This paper is organized as follows.
Section 2 reviews
5D SUGRA in $\scr{N}=1$ superspace \cite{LLP},
and proves the existence of the remarkably simple gauge fixing noted above.
Section 3 gives the supergraph Feynman rules for the theory.
Sections 4 and 5 contain the calculations of $c_1$ and $c_2$, respectively.  We find that both $c_1$ and $c_2$ give negative scalar mass-squared terms in the visible sector.  The result for $c_1$ agrees with \Ref{GR}, while the result for $c_2$ is new.
\section{Lagrangian and Gauge-fixing}
The Lagrangian for linearized minimal 5D SUGRA was written in
terms of $\scr{N} = 1$ superfields in \Ref{LLP}.
Here, we describe the component field embedding and state the superfield
action.  We then prove the existence of the gauge choice
\Eq{boney}.

The formulation of \Ref{LLP}
contains two real superfields $V_m$ and $P$, a chiral superfield
$T$,
and an unconstrained superfield $\Psi_\al$.%
\footnote{The field $\Psi_\al$ corresponds to what was called
$\hat{\Psi}_\al$ in \Ref{LLP}.}
%
The embedding of the 5D propagating fields into these superfields
is accomplished as follows.
The graviton, graviphoton and gravitino are first dimensionally reduced:
\beq\bal
h_{MN} &\to h_{mn},\ h_{5m}, h_{55},
\\
B_M &\to B_m, B_5,
\\
\psi_{M\tilde{\al}} &\to \psi_{m\al}^{(\pm)},
\eal
\eeq
Here the 5D gravitino is decomposed into
components with parity $\pm 1$ under the $Z_2$ transformation
$x^5 \mapsto -x^5$.
These reduced fields are embedded in superfields as
\beq\bal
V_m &\sim \theta \si^n \bar\theta h_{mn} +
{\bar\theta}^2 \theta^\al \psi_{m \al}^{(+)} + \cdots,
\\
\Psi_\al &\sim \bar{\theta}^{\dot\al} (B_{\al\dot\al}
+ i h_{5 \al\dot\al}) + \theta \si^m \bar\theta \psi_{m \al}^{(-)}
+ \bar\theta^2 \psi_{5 \al}^{(-)} + \cdots,
\\
T &\sim h_{55} + i B_5 + \theta^\al \psi_{5 \al}^{(+)} + \cdots.
\eal\eeq
In this formulation, when the $Z_2$ even superfields
$V_m$ and $P$ are evaluated on either boundary they are the usual
4D $\scr{N} = 1$ SUGRA multiplet.
(The real field $P$ is the
prepotential for the usual conformal compensator: $\Si = -\sfrac
14 \bar{D}^2 P$.)
This makes coupling 5D SUGRA to fields localized
on the boundaries particularly simple. For details, see \Ref{LLP}.

The Lagrangian for linearized 5D SUGRA is
\beq[Lsugra]
\scr{L}_{\rm 5D\, SUGRA} = \scr{L}_{\scr{N} = 1} + \De \scr{L}_{5},
\eeq
where $\scr{L}_{\scr{N} = 1}$ is the linearized $\scr{N} = 1$ SUGRA Lagrangian
(see e.g. \cite{BK})%
\footnote{We use the conventions of Wess and Bagger \cite{WB}.}
\beq[LN1]
\bal
\scr{L}_{{\cal N} = 1} &=
M_5^3 \myint d^4\theta \Bigr[
\sfrac 18 V^m D^\al \bar{D}^2 D_\al V_m
+ \sfrac 1{48} \left( [D^{\al}, \bar{D}^{\dot\al} ] V_{\al\dot\al} \right)^2
- (\d^m V_m)^2
\\
& \qquad\qquad\qquad
- \sfrac 13 \Si^\dagger \Si
+ \sfrac {2i}3 (\Si - \Si^\dagger) \d^m V_m \Bigr],
\eal
\eeq
and
\beq
[L5] \bal \De{\scr{L}}_5 = -\sfrac 12 M_5^3 \myint
d^4\theta \Bigl\{ & \bigl[  T^\dagger ( \Si - i \d_{\al \dot\al}
V^{\dot\al \al}) + \hc \bigr] \\ & - \sfrac 12 \bigl[ D^\al \Psi_\al +
\bar{D}_{\dot\al} \Psi^{\dagger \dot\al} - \d_5 P \bigr]^2 \\ & +
\bigl[ \d_5 V_{\al\dot\al} - (\bar{D}_{\dot\al} \Psi_\al - D_\al
\Psi^\dagger_{\dot\al} ) \bigr]^2 \Bigr\}.
\eal
\eeq
In this
normalization, $M_{\rm P}^2 = \frac 12 \pi r M_5^3$, where
$M_{\rm P} = 2 \times 10^{18}\GeV$ is the 4D Planck scale.

The terms in the Lagrangian involving the brane-localized superfields $X$ and $Q$ are
\beq
\De \scr{L}_{\rm brane} = \de(x^5) \scr{L}_{4,{\rm kin}}(Q)
+ \de(x^5 - \ell) \scr{L}_{4,{\rm kin}}(X),
\eeq
where $\scr{L}_{4,{\rm kin}}(\Phi)$ is the kinetic term for a 4D chiral superfield $\Phi$ coupled to 4D SUGRA:
\beq[vert]
\scr{L}_{4,{\rm kin}}(\Phi) = \myint d^4\theta \left[
\Phi^\dagger \Phi + \sfrac{2i}{3} V^m \Phi^\dagger \dd_m \Phi
- \sfrac 16 V^m K_{mn} V^n \Phi^\dagger \Phi + \cdots \right].
\eeq
Here we have absorbed the conformal compensator $\Si$ into $\Phi$.
We have omitted terms $\scr{O}(V^3)$ and higher, as well as $\scr{O}(V^2)$
with derivatives acting on the chiral fields, since these do not contribute
to the terms in \Eq{c12}.
Finally, $K_{mn}$ represents the quadratic terms in \Eq{LN1}
and is given explicitly by:
\bea
K_{mn}:=
\sfrac 14 \eta_{mn}D^\al\bar D^2D_\al
+\sfrac 1{24} \si_m^{\dot\al \al} \si_n^{\dot\be \be}
[D_\al,\bar D_{\dot\al}][D_\be,\bar D_{\dot\be}]
+2\partial_m\partial_n
\eea

To define the propagator for quantum calcuations, we must first fix the gauge.
We require just the $V_mV_n$ propagator,
because the vertices from \Eq{vert} involve only $V_m$.  We now show
that there exists a gauge fixing term that cancels the mixing
between $V_m$ and the other bulk superfields $P$, $\Psi_\al$, $T$,
and simultaneously reduces the $V_m$
kinetic term to the simplest possible form $V^m \Box_5 V_m$.
To do this, we rewrite the quadratic terms in $V$ as
\beq
\scr{L}_{\rm 5D\,SUGRA} = M_5^3 \myint d^4\theta \left[
-\sfrac 12 V^m \Box_5 V_m
+ \scr{Q}(\bar{D}_{\dot\al} V^{\dot\al \al}) + \cdots \right],
\eeq
where
\beq \scr{Q}(\chi^\al) = \sfrac 1{24} \chi^2 - \sfrac
14 \chi^\al (\bar{D}_{\dot \al} D_\al - \sfrac 13 D_\al
\bar{D}_{\dot\al}) \bar{\chi}^{\dot\al}.
\eeq
We then add the gauge fixing term \beq \De \scr{L}_{\rm gf} =
-M_5^3 \myint d^4 \theta \, \scr{Q}(\scr{G}^\al), \eeq where the
gauge fixing function takes the form \beq\bal \scr{G}^\al &=
\bar{D}_{\dot\al} V^{\dot\al \al} + \frac{\d_5}{\Box_4} \left(
\bar{D}^2 \Psi^\al - \sfrac{6i}{5} \partial^{\dot\al \al}
\bar{\Psi}_{\dot\al} - \sfrac 25 [D^\al, \bar{D}^{\dot\al}]
\bar{\Psi}_{\dot\al} \right) \\ & \qquad - \sfrac i5
\partial^{\dot\al \al} \bar{D}_{\dot\al} (T^\dagger - \sfrac 13
\Si^\dagger).  \eal \eeq With this addition, we have \beq
\scr{L}_{\rm 5D\,SUGRA} +\De \scr{L}_{\rm gf}= M_5^3 \myint
d^4\theta\, \left[ -\sfrac 12 V^m \Box_5 V_m \right]+ \scr{L}( P, T,
\Psi_\al). \eeq Note that we do not need the ghost action, since
we are not computing loops involving SUGRA self-couplings. Hence
the ghosts decouple and do not
contribute to the quantities under consideration.
%

\section{Superpropagators on the Orbifold}
The perturbative theory for the model with the superfield
Lagrangian given by \Eq{Lsugra} can be completely formulated in
terms of superfields with the help of supergraphs (see e.g.
\cite{GGRS}, \cite{BK}).
Although we will not review these
techniques, we will describe the relevant modifications to
describe the brane-world scenario.
In this brane-world scenario we have an $S^1 / Z_2$ orbifold.
Thus, it is convenient to write
the Feynman rules in mixed 4D momentum space and 5D position
space.
Further, for supergraph calculations, we use the following
abbreviations
\beq
\int_{1,\ldots,n} = \myint d^4\theta_1 \cdots
\myint d^4\theta_n,
\quad
\de_{12} = \de^4(\theta_1 - \theta_2),
\quad
\scr{D}_1(p) = -\sfrac 14 D^2_1(p),
\eeq
where $D_\al(p)$ is the SUSY covariant derivative in momentum space.
We omit the $p$ argument when this leads to no ambiguity.
We also note the following identities used for manipulating covariant
derivatives and delta functions under superspace integrals:
\beq[deeznuts]\bal
{\cal D}_1 \de_{12} &= {\cal D}_2\de_{12} ,\cr
\de_{12}({\cal D}_1\bar {\cal D}_1\de_{12})
&= \de_{12}(\bar {\cal D}_1 {\cal D}_1\de_{12}) = \de_{12},\cr
\de_{12} \bigl[ {\cal O}(D_1^n\bar D^m_1)\de_{12} \bigr]
&= 0\ {\rm for}\ n < 2\ {\rm or}\ m < 2 ,\cr
\de_{12}({\cal D}_1\bar {\cal D}_1{\cal D}_1\bar {\cal D}_1\de_{12})
&=\de_{12}(\bar{\cal D}_1 {\cal D}_1\bar{\cal D}_1{\cal D}_1\de_{12})
=-p^2\de_{12}.
\eal\eeq

The $V_m$ propagator with one endpoint fixed on the visible brane
is
\beq[corn]
\avg{V_m(1, x^5 = 0) V_n(2, x^5 = y)} = i
\eta_{mn}\delta_{12} \De (p, y),
\eeq where the Green function
$\De(p,y)$ satisfies the equation
\beq
M_5^3(\d_y^2 - p^2) \De(p, y) =- \de(y).
\eeq
Since $V_m$ is an even field, the boundary
conditions are $\d_y \De = 0$ at the branes.
In the domain $-\ell < y < \ell$, we have
\beq
\De(p, y) = \frac{1}{M_5^3} \,
\frac{\cosh(p(|y| - \ell))}{2p\sinh(p\ell)},
\eeq
where
$p = +\sqrt{p^m p_m}$.
The propagator from the visible brane to the
visible brane is given by the limit $y \to 0$:
\beq[visvis]
\De_{\rm vis,vis}(p) = \frac{1}{M_5^3}\, \frac{1}{2 p \tanh(p\ell)}.
\eeq
The propagator between the visible and hidden
branes is given by the limit $y \to \ell$:
\beq[vishid]
\De_{\rm vis,hid}(p) = \frac{1}{M_5^3} \frac{1}{2p \sinh(p \ell)}.
\eeq
The chiral propagators localized on the brane are given by the
standard 4D expression (see e.g. \cite{GGRS}, \cite{BK})
\beq[chips]
\avg{\Phi^\dagger(1) \Phi(2)} = -\frac{i}{p^2}
\scr{D}_1 \bar{\scr{D}}_2 \de_{12}.
\eeq
The vertices between chiral fields and $V_m$ are read off from \Eq{vert}.

We now have all of the necessary ingredients to compute the
coefficients $c_1$ and $c_2$.  We neglect contributions due to the derivatives of $Q, Q^\dagger$ and $X, X^\dagger$, because we are only interested in corrections to scalar masses.
Since  $c_1$ and $c_2$ are gauge invariant, they can be computed using the
simple gauge choice described above.

A few comments about the supergraph technique are in order.  The standard procedure of
supergraph calculations is reviewed in \Refs{GGRS} and \cite{BK}.
The main feature is that SUSY is manifest at every step.  Another feature of the supergraph technique is the presence of SUSY covariant derivatives and Grassman $\de$-functions.  The covariant derivative algebra is what simplifies the calculations
in comparison to component formulations of SUSY theories.
In an arbitrary supergraph, one can transfer all covariant derivatives onto one
Grassman $\de$-function using integration by parts.
This removes all integrals over anticommuting variables except one.
Then one can transform the last integral over superspace to a standard
Feynman integral over conventional momentum space.
This is accomplished by using the rules given in \Eq{deeznuts}.
This procedure avoids calculating large numbers of diagrams that would appear in a component formulation of a SUSY theory.
Furthermore, it automatically accounts
the cancelations of conventional diagrams stipulated by $\scr{N} = 1$ SUSY.

\section{Radion-mediated Contribution}
We now compute the coefficient $c_1$ in the effective lagrangian \Eq{c12}.
The 1-loop supergraphs that contribute are shown in Fig.~1.

\begin{figure}[t]
\begin{center}
\begin{picture}(320,100)(0,0)
 \Line(-50,5)(140,5)
 \Line(-50,5)(-40,35)
 \Line(140,5)(130,35)
 \Line(130,35)(-40,35)
 \ArrowLine(125,15)(-35,15)
 \Text(105,25)[c]{$\theta_1$}
 \Text(-10,25)[c]{$\theta_2$}
 \Text(45,25)[c]{$p-k$}
 \DashCArc(45,15)(70,0,180){1}
 \Text(45,65)[c]{$k$}
\LongArrowArc(45,15)(65,75,105)
\Vertex(-25,15){1}
\Vertex(115,15){1} \Text(45,-5)[c]{$(a)$}
\Line(171,5)(361,5)
\Line(171,5)(181,35)
\Line(361,5)(351,35)
\Line(351,35)(181,35)
\ArrowLine(346,15)(266,15)
\ArrowLine(266,15)(186,15)
\Text(266,25)[c]{$\theta_1$}
\DashCArc(266,55)(40,-90,270){1}
\Text(266,83)[c]{$k$}
\LongArrowArc(266,55)(35,60,120)
\Vertex(266,15){1}
\Text(266,-5)[c]{$(b)$}
\end{picture}
\caption{\label{FeynD1}Supergraphs contributing to the coefficient
$c_1$, the radion mediated corrections to SUSY breaking. }
\end{center}
\end{figure}
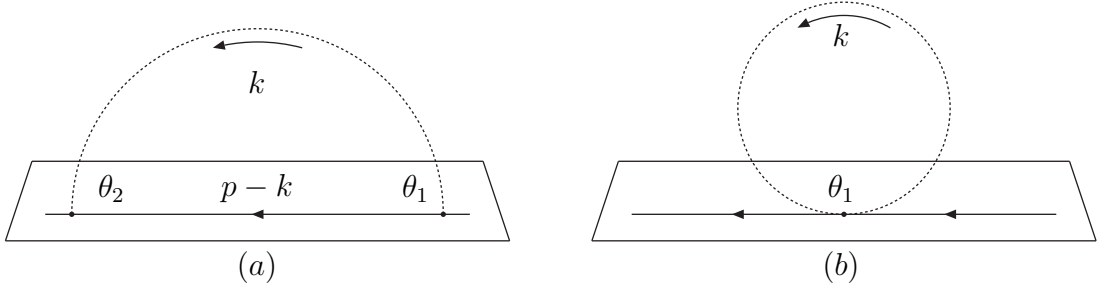

The diagram in Fig.~1a is given by
\beq \hbox{\rm Fig.~1a} = -\left( \sfrac 23 \right)^2 \myint
\frac{d^4 p}{(2\pi)^4} \De_{\rm vis,vis}(p) \frac{p_m p_n}{p^2} \,
I^{mn}_{\rm 1a}, \eeq
The superspace integral $I^{mn}_{\rm 1a}$ is
\beq[1a]
I^{mn}_{\rm 1a} = \eta^{mn}
\int_{1,2} \de_{12} \scr{D}_1 \bar{\scr{D}}_2 \de_{12} =
\eta^{mn} \int_1,
\eeq
and leads to
\beq[c11a] \hbox{\rm Fig.~1a}
= -\left( \sfrac 23 \right)^2 \myint d^4\theta \myint \frac{d^4
p}{(2\pi)^4} \De_{\rm vis,vis}(p).
\eeq
The second diagram Fig.~1b gives
\beq \hbox{\rm
Fig.~1b} = \sfrac 16\myint \frac{d^4 p}{(2\pi)^4} \De_{\rm
vis,vis}(p) \, I_{\rm 1b}.
\eeq
For this diagram, the $V_m$ propagator must be evaluated in the limit that $\theta_1$ goes to $\theta_2$.  This is equivalent to inserting one more delta-function and
integrating over $\theta_2$.  Thus, the superspace integral
$I_{\rm 1b}$ is
\beq[1b]
I_{\rm 1b} = \int_{1,2} \de_{12}\eta^{mn} K_{mn} \de_{12} = \sfrac {32}{3} \int_1,
\eeq
which yields
\beq[c11b]
\hbox{\rm Fig.~1b} = \sfrac{16}{9} \myint d^4\theta \myint \frac{d^4
p}{(2\pi)^4} \De_{\rm vis,vis}(p).
\eeq

The momentum integral is UV divergent, but its divergent part is
independent of $\ell$.
This is easily seen from the leading behavior
of the propagator at large $p$, which is $\De \to 1/(2p)$.
Physically, this UV
divergent contribution renormalizes the $Q$ kinetic term on the visible
brane, which is insensitive to the size of the extra dimension.
For radion-mediated SUSY breaking we are interested in the $\ell$
dependent contribution, so we write
\beq[c1pint]
\myint \frac{d^4 p}{(2\pi)^4} \De_{\rm vis,vis}(p)
&= \frac{1}{M_5^3}\myint \frac{d^4 p}{(2\pi)^4} \,
\frac{1}{2 p \tanh(p\ell)}
\nonumber\\
&= \frac{1}{M_5^3}\myint \frac{d^4 p}{(2\pi)^4} \,
\frac{e^{-p\ell}}{2 p \sinh(p\ell)}
+ \hbox{\rm independent\ of\ }\ell
\nonumber\\
&= \frac{1}{4\pi^2 M_5^3}\frac{\zeta(3)}{(2\ell)^3}
+ \hbox{\rm independent\ of\ }\ell.
\eeq
where $\zeta(3) \simeq 1.202$ is the Riemann zeta function.
Combining \Eqs{c11a}, \eq{c11b}, and \eq{c1pint} the total result for $c_1$ is
\beq[c1]
c_1 = \frac{1}{3\pi^2 M_5^3}\frac{\zeta(3)}{(2\ell)^3}.
\eeq

\section{Brane-to-Brane Contribution}
We now compute the coefficient $c_2$ in the effective lagrangian \Eq{c12}.
The 1-loop supergraphs that contribute are shown in Fig.~2.
We first consider the diagram of Fig.~2a, consisting of four
4-point interactions.
There are two possible contractions for this
diagram.
One of them vanishes due to the SUSY covariant derivative algebra, and
the other yields
\beq \hbox{Fig.~2a} = \left( \sfrac 23 \right)^4
\myint \frac{d^4 p}{(2\pi)^4} \left[ \De_{\rm vis,hid}(p)
\right]^2 I_{\rm 2a} .
\eeq
The superspace integral $I_{\rm 2a}$ is
\begin{figure}[t]
\begin{center}
\begin{picture}(320,100)(0,0)
 \Line(-50,10)(-50,90)
 \Line(-50,10)(-28,20)
 \Line(-28,80)(-50,90)
 \Line(-28,80)(-28,20)
\Line(66,10)(66,90)
\Line(66,10)(44,20)
\Line(44,80)(66,90)
\Line(44,80)(44,20)
\Line(-38,20)(-38,80)
\Line(54,20)(54,80)
\DashLine(-38,30)(54,30){1}
\DashLine(-38,70)(54,70){1}
\Vertex(-38,30){1}
\Vertex(-38,70){1}
\Vertex(54,30){1}
\Vertex(54,70){1}
\LongArrow(-5,25)(25,25)
\LongArrow(25,75)(-5,75)
\LongArrow(-42,63)(-42,37)
\LongArrow(58,37)(58,63)
\Text(-44,75)[c]{$\theta_1$}
\Text(60,75)[c]{$\theta_2$}
\Text(-44,25)[c]{$\theta_4$}
\Text(60,25)[c]{$\theta_3$}
\Text(10,85)[c]{$k$}
\Text(10,15)[c]{$k$}
\Text(-46,50)[c]{$k$}
\Text(62,50)[c]{$k$}
\Text(10,0)[c]{$(a)$}
 \Line(101,10)(101,90)
 \Line(101,10)(122,20)
 \Line(122,80)(101,90)
 \Line(122,80)(122,20)
\Line(213,10)(213,90)
\Line(213,10)(192,20)
\Line(192,80)(213,90)
\Line(192,80)(192,20)
\Line(112,20)(112,80) \Line(202,20)(202,80)
\DashCArc(157,-4)(70,51,129){1} \DashCArc(157,104)(70,231,309){1}
\Vertex(112,50){1} \Vertex(202,50){1}
\LongArrowArc(157,2)(70,75,105)
\LongArrowArc(157,98)(70,255,285)
\Text(108,50)[c]{$\theta_1$} \Text(209,50)[c]{$\theta_2$}
\Text(157,80)[c]{$k$} \Text(157,20)[c]{$k$} \Text(157,0)[c]{$(b)$}
\Line(248,10)(248,90)
\Line(248,10)(270,20)
\Line(270,80)(248,90)
\Line(270,80)(270,20)
\Line(361,10)(361,90)
\Line(361,10)(340,20)
\Line(340,80)(361,90)
\Line(340,80)(340,20)
\Line(259,20)(259,80)
\Line(350,20)(350,80)
\DashLine(259,30)(350,50){1}
\DashLine(259,70)(350,50){1}
\Vertex(259,30){1}
\Vertex(259,70){1}
\Vertex(350,50){1}
\LongArrow(319,63)(290,70)
\LongArrow(288,28)(317,35)
\LongArrow(256,63)(256,37)
\Text(256,76)[c]{$\theta_1$}
\Text(357,50)[c]{$\theta_2$}
\Text(255,25)[c]{$\theta_3$}
\Text(306,75)[c]{$k$}
\Text(306,25)[c]{$k$}
\Text(253,50)[c]{$k$}
\Text(306,0)[c]{$(c)$}
\end{picture}
\caption{\label{FeynD2}Supergraphs contributing to the coefficient
$c_2$, the brane-to-brane corrections to SUSY breaking. }
\end{center}
\end{figure}
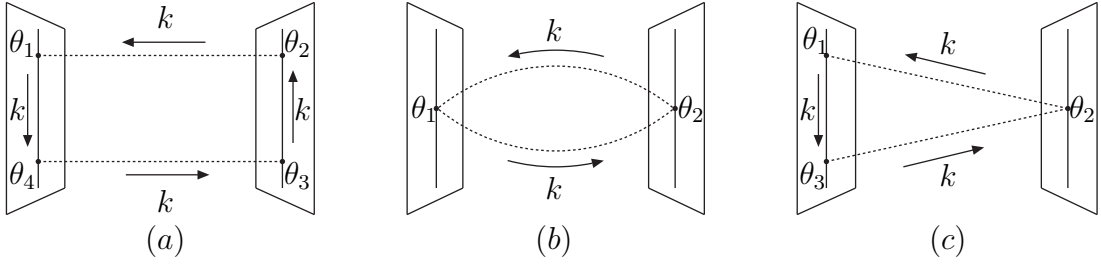
\beq[2a]
I_{\rm 2a} &= \int_{1,\ldots,4} \left( \scr{D}_1 \bar{\scr{D}_3} \de_{13} \right)
\left( \scr{D}_4 \bar{\scr{D}_2} \de_{24} \right) \de_{12} \de_{34}
= -\int_1 p^2,
\eeq
and gives
\beq[2afin]
\hbox{\rm Fig.~2a} =
-\left( \sfrac 23 \right)^4 \myint d^4\theta \myint \frac{d^4
p}{(2\pi)^4} \, p^2 \left[ \De_{\rm vis,hid}(p) \right]^2 .
\eeq
The diagram of Fig.~2b contains two 4-point interactions
\beq
\hbox{\rm Fig.~2b} &= 2 \left( \sfrac 16
\right)^2 \myint \frac{d^4 p}{(2\pi)^4} \left[ \De_{\rm
vis,hid}(p) \right]^2 I_{\rm 2b} ,
\eeq
where the superspace
integral $I_{\rm 2b}$ is
\beq[2b]
I_{\rm 2b}
= \int_{1,2} \de_{12} \left(K^{mn}_1 K_{nm,1} \de_{12} \right)
= \sfrac{112}{9} \int_1 p^2,
\eeq
and leads to
\beq[2bfin] \hbox{\rm Fig.~2b} &= \sfrac{224}{9}\left( \sfrac 16
\right)^2 \myint d^4 \theta \myint \frac{d^4 p}{(2\pi)^4} \left[
\De_{\rm vis,hid}(p) \right]^2 p^2.
\eeq
The diagram Fig.~2c contains two 3-point and one 4-point interaction.
There are two contractions, each giving the same contribution.
We obtain
\beq \hbox{\rm Fig.~2c} = 2 \times (-\sfrac 13)
\left( \sfrac 23 \right)^2 \myint \frac{d^4 p}{(2\pi)^4}\,
\frac{p_m p_n}{p^2} \left[ \De_{\rm vis,hid}(p) \right]^2
I^{mn}_{\rm 2c} ,
\eeq
where the superspace integral $I^{mn}_{\rm 2c}$ is
\beq[2c]
I^{mn}_{\rm 2c} = \int_{1,2,3} \left( \scr{D}_1 \bar{\scr{D}}_2 \de_{12} \right)
\left( K^{mn}_3 \de_{13} \right) \de_{23}
= 
-\sfrac{4}{3}\int_1 p^m p^n ,
\eeq
and yields
\beq[2cfin] \hbox{\rm
Fig.~2c} = \sfrac 89 \left( \sfrac 23 \right)^2 \myint \frac{d^4
p}{(2\pi)^4}\, p^2 \left[ \De_{\rm vis,hid}(p) \right]^2.
\eeq

The momentum integral is UV finite and can be evaluated directly.
Physically, the UV finiteness is due to the fact that the SUGRA
propagator cannot shrink to zero size because the endpoints are fixed
on different branes.
The integral is
\beq[2pint]
 \myint \frac{d^4 p}{(2\pi)^4}\, p^2 \left[ \De_{\rm vis,hid}(p) \right]^2
&=\frac{1}{4M_5^6}
\myint \frac{d^4 p}{(2\pi)^4}\, \frac{1}{ \sinh^2(p \ell)}
\cr
& = \frac{3}{4\pi^2M_5^6}\frac{\zeta(3)}{(2\ell)^4}
\eeq
Combining \Eqs{2afin}, \eq{2bfin}, \eq{2cfin}, and \eq{2pint} the final result is
\beq[c2]
c_2 = \frac{2}{3\pi^2M_5^6}\frac{\zeta(3)}{(2\ell)^4}.
\eeq

This completes our calculation.
The coefficients $c_1$ and $c_2$ in the 4D effective lagrangian
defined in \Eq{c12} are given by \Eqs{c1} and \eq{c2}, respectively.

\section{Conclusion}
We have formulated an $\scr{N} = 1$ supergraph approach to 5D supergravity
(SUGRA)
loop calculations, using the formulation of 5D SUGRA in terms of
$\scr{N} = 1$ superfields of \Ref{LLP}.
This formalism makes $\scr{N} = 1$ SUSY manifest, and makes couplings
between bulk and brane fields particularly simple.
In particular, there are no terms with higher powers of delta functions
appearing in the calculation, as in component approaches.

We applied this formalism to compute the leading SUGRA loop contributions to
visible sector scalar masses in the simplest `brane world' scenario based
on a flat 5D space compactified on a $S^1 / Z_2$ orbifold.
The terms in the effective lagrangian are defined in \Eq{c12} and our results
are given in \Eqs{c1} and \eq{c2}.
The calculation requires the calculation of only five supergraphs.

The same effective lagrangian terms have been calculated by
R. Rattazzi, C.A. Scrucca, and A. Strumia using the component formulation of 5D supergravity.
Our results agree \cite{RSS}.

There are a number of directions to extend the present results.
Warped compactifications may give positive loop contributions to scalar masses.
It would also be interesting to extend the present results to higher dimensions,
possibly to make direct contact with string theory, and also to construct
the fully nonlinear theory.
We leave these questions to future work.

\section*{Acknowledgements}
Work of I.L.B was supported by INTAS grant project 991-590, RFBR grant project 03-02-16193, joint RFBR-DFG grant project 02-02-04002, DFG grant project 436RUS 113/669 and grant for Leading Russian Scientific Schools project 1252.2003.2.  S.J. Gates, H.S. Goh, M.A. Luty, and S.P.~Ng were supported by NSF grant PHY-0099544.  S.J.~Gates, W.D.~Linch III, and J.~Phillips were supported by the University of Maryland Center for String and Particle Theory (CSPT). I.L.B. was also partially supported by the CSPT.


\newpage
\begin{thebibliography}{99}
\bibitem{RS}
L.~Randall and R.~Sundrum,
Nucl.\ Phys.\ B {\bf 557}, 79 (1999)\\
\href{http://arXiv.org/abs/hep-th/9810155}{[arXiv:hep-th/9810155]}.
\bibitem{GLMR}
G.~F.~Giudice, M.~A.~Luty, H.~Murayama and R.~Rattazzi,
JHEP {\bf 9812}, 027 (1998)
\href{http://arXiv.org/abs/hep-ph/9810442}{[arXiv:hep-ph/9810442]}.
\bibitem{LS1}
M.~A.~Luty and R.~Sundrum, Phys.Rev.D {\bf 62}:035008, 2000\\
 \href{http://arXiv.org/abs/hep-th/9910202}{[arXiv:hep-th/9910202]}.
\bibitem{SS}
J.~Scherk and J.~H.~Schwarz,
Phys.\ Lett.\ B {\bf 82}, 60 (1979)\\ \href{http://ccdb3fs.kek.jp/cgi-bin/img_index?7901033}{[KEK Library: 7901033]}.
\bibitem{SSequiv}
D.~Marti and A.~Pomarol,
Phys.\ Rev.\ D {\bf 64}, 105025 (2001)
\href{http://arXiv.org/abs/hep-th/0106256}{[arXiv:hep-th/0106256]};
J.~A.~Bagger, F.~Feruglio and F.~Zwirner,
Phys.\ Rev.\ Lett.\  {\bf 88}, 101601 (2002)
\href{http://arXiv.org/abs/hep-th/0107128}{[arXiv:hep-th/0107128]};
D.~E.~Kaplan and N.~Weiner,
\href{http://arXiv.org/abs/hep-ph/0108001}{[arXiv:hep-ph/0108001]}.
\bibitem{GR}
T.~Gherghetta and A.~Riotto,
Nucl.\ Phys.\ B {\bf 623}, 97 (2002)\\
\href{http://arXiv.org/abs/hep-th/0110022}{[arXiv:hep-th/0110022]}.
\bibitem{Zucker}
M.~Zucker,
Nucl.\ Phys.\ B {\bf 570}, 267 (2000)
\href{http://arXiv.org/abs/hep-th/9907082}{[arXiv:hep-th/9907082]}.
\bibitem{GGRS}
S.J. Gates, M.T. Grisaru, M. Ro\v cek, W. Siegel, {\it Superspace},
Benjamin Cummings, Reading, MA, 1983 \href{http://arXiv.org/abs/hep-th/0108200}{[arXiv:hep-th/0108200]}.
\bibitem{BK}
I.L. Buchbinder, S.M. Kuzenko, {\it Ideas and Methods of
Supersymmetry and Supergravity}, IOP Publ., Bristol and Philadelphia,
1998.
\bibitem{LLP}
W.~D.~Linch, M.~A.~Luty and J.~Phillips,
\href{http://arXiv.org/abs/hep-th/0209060}{arXiv:hep-th/0209060},
to be published in Phys.\ Rev.\ D.
\bibitem{WB}
J. Wess and J. Bagger, {\it Supersymmetry and Supergravity},
Princeton Univ.Press, Princeton, 1992.
\bibitem{MP}
E.~A.~Mirabelli and M.~E.~Peskin,
Phys.\ Rev.\ D {\bf 58}, 065002 (1998)\\
\href{http://arXiv.org/abs/hep-th/9712214}{[arXiv:hep-th/9712214]}.
\bibitem{RSS}
R. Rattazzi, C.A. Scrucca, A. Strumia, private communication.
\end{thebibliography}
\end{document}